\documentclass[twocolumn]{article}  

    \usepackage[bookmarksnumbered,pdfpagelabels=true,plainpages=false,colorlinks=true,linkcolor=blue,citecolor=red,urlcolor=blue]{hyperref}
	\usepackage{graphicx}
	\graphicspath{{pictures/}}
	\usepackage[labelformat=simple]{subcaption} 
	\usepackage[belowskip=-28pt]{caption}   
	\usepackage{dblfloatfix} 
	\usepackage{dcolumn}   
	\usepackage{bm}        
	\usepackage{amssymb}   
	\usepackage{amsmath}   
	\usepackage{cleveref}  
	\usepackage{siunitx}   
	\usepackage{IEEEtrantools} 
	\usepackage{xcolor}
	\usepackage[]{authblk} 
	\usepackage{siunitx}

\title{\textbf{Magnetization dynamics in synthetic antiferromagnets: the role of dynamical energy and mutual spin-pumping}}
\author[1,2]{S. Sorokin}
\author[3,4]{R. Gallardo}
\author[1]{C. Fowley}
\author[1]{K. Lenz}
\author[1]{A. Titova}
\author[5]{G. Atcheson}
\author[5]{G. Dennehy}
\author[5]{K. Rode}
\author[1,2]{J. Fassbender}
\author[1]{J. Lindner}
\author[1]{A. M. Deac}
\affil[1]{Institute of Ion Beam Physics and Materials Research, Helmholtz - Zentrum Dresden - Rossendorf, Bautzner Landstr. 400, 01328, Dresden, Germany}
\affil[2]{Institute for Physics of Solids, Technische Universität Dresden, Dresden, Germany }
\affil[3]{Departamento de F\'{i}sica, Universidad T\'{e}cnica Federico Santa Mar\'{i}a, Avenida Espa\~{n}a 1680, Valpara\'{i}so, Chile}
\affil[4]{Center for the Development of Nanoscience and Nanotechnology (CEDENNA), 917-0124 Santiago, Chile}
\affil[5]{CRANN and School of Physics, Trinity College Dublin, Dublin 2, Ireland}

\date{}	
\begin{document}
	
\raggedbottom
	\twocolumn[
	\begin{@twocolumnfalse}
		\maketitle
		\begin{abstract}
		We investigate magnetization dynamics in asymmetric interlayer exchange coupled Py/Ru/Py trilayers using both vector network analyzer-based and electrically-detected ferromagnetic resonance techniques. 
		Two different ferromagnetic resonance modes, in-phase and out-of-phase, are observed across all three regimes of the static magnetization configurations, through antiparallel alignment at low fields, the spin-flop transition at intermediate fields and the parallel alignment at high fields. 
		The non-monotonic behavior of the modes as a function of the external field is explained in detail by analyzing the interlayer exchange and Zeeman energies, and is found to be solely governed by the interplay of their dynamical components. In addition, the linewidths of both modes were determined across the three regimes and the  different behaviors of the linewidths versus external magnetic field are attributed to mutual spin pumping induced in the samples. Interestingly, the difference between the linewidths of the out-of-phase and in-phase modes decreases at the spin-flop transition and is reversed between the antiparallel and parallel aligned magnetization states.
		\end{abstract}
	\end{@twocolumnfalse}
\vspace{20pt}
	]

\section{Introduction}

Exchange coupled ferro-, ferri-, and antiferromagnetic multilayers are extensively used in magnetic storage devices \cite{Mot_1}, magnetic read-heads \cite{Mot_2}, non-volatile magnetic random access memory (MRAM) \cite{Mot_3} and spin-torque oscillators \cite{SAF_oscillator,SAF_oscillator_exp,SAF_oscillator_th,SAF_AF_oscillator_exp}. 
The base frequency of such oscillators can be tuned in the range from several to tens of GHz by varying composition and geometry \cite{Mot_4,Mot_5,Lorenzo_IEC}. 
A synthetic antiferromagnet (SAF) is a simple trilayer structure consisting of two thin ferromagnetic films seperated by a thin spacer through which the magnetizations are coupled by interlayer exchange coupling (IEC) \cite{Parkin_IEC, Fert_GMR, Grunberg_GMR}.
Negative IEC promotes antiparallel alignment of the two layers, competing with the Zeeman energy, that tends to align both magnetizations along the external field.
Three regimes can be identified, depending on the magnitude of the applied field:
the \emph{antiferromagnetically (AF) coupled} regime where the IEC dominates at low magnetic fields; \textit{the Zeeman dominated} regime where both magnetizations are \emph{saturated} along the applied field direction; and the \emph{spin-flop regime}, which separates the former two. 
The existence of these distinct regimes result in a non-monotonic behavior of the frequency as a function of the field [$f(\mathbf{H})$] \cite{Spin-Flop_ex1,Spin-Flop_ex2,Rezende98}. 
So far, dynamical studies on SAFs were mostly performed using conventional (cavity) ferromagnetic resonance (FMR) \cite{Zhang_Long_FMR} and vector network analyzer-based ferromagnetic resonance (VNA-FMR) \cite{Belmequenai_2007,Belmequenai_2008} approaches. 
This allowed to identify the two eigenmodes, usually referred to as ``acoustic'' -- for the in-phase precessing magnetizations and ``optic'' -- for the out-of-phase precession of the magnetizations of the two layers \cite{Lorenzo_IEC,Zhang_Long_FMR,Belmequenai_2007,Belmequenai_2008}. 
Up to now, however, the non-monotonic behavior $f(\mathbf{H})$ of the modes has not been properly explained in the AF-coupled regime  \cite{Belmequenai_2008,Py_Ru_Py_SAF_Dyn_notExpl,Py_Ru_Py_Pt_SAF_Dyn_notExpl}. 
In this regime $f(\mathbf{H})$, the acoustic (optic) mode exhibits a distinct maximum (minimum), despite of no change in the static magnetic configuration \cite{Belmequenai_2008,Py_Ru_Py_SAF_Dyn_notExpl,Py_Ru_Py_Pt_SAF_Dyn_notExpl}.
Here, we show that $f(\mathbf{H})$ in asymmetric Py/Ru/Py SAFs depends on the interplay between the dynamical components of the Zeeman and IEC energy terms, and therefore not  on the static configurations specifically.

In addition, in such SAF systems the effect of mutual spin-pumping was predicted to influence the linewidth of both modes \cite{Takanashi,IEC_SP_theory}.
The effect was observed in symmetric Py/Ru/Py trilayers but only when both layers were saturated at high fields, beyond the spin-flop \cite{Yang_2016}. 
We present the experimental observation of mutual spin-pumping across a wide field range, covering the antiferomagnetically coupled and saturated alignment, as well as through the spin-flop regime, where the linewidth difference reverses sign. 

This paper is organised as follows:
First we introduce the material system we used and describe the preliminary magnetic and electrical characterisations and determine the expressions for the IEC and Zeeman energies.  
Magnetization dynamics of the same structures and a model explaining the behavior is then presented.
The model is capable of predicting both the frequency dependence and amplitude of the measured VNA-FMR signal, and shows that the behavior is a result of the interplay not only of the static, but also the dynamical components of the Zeeman and IEC energies.
Finally the change in linewidth due to mutual spin-pumping between the layers is evaluated and a connection between the non-monotonic $f(\mathbf{H})$ dependence and the variation of the linewidth difference over the full field range is established.


\section{Methodology and Results}
\subsection{Static magnetic states in asymmetric SAFs}

\begin{figure}[]	
	\begin{subfigure}[!t]{\linewidth}
		\captionsetup{labelfont=bf,font=large}
		\caption{}
		\includegraphics[width=\linewidth]{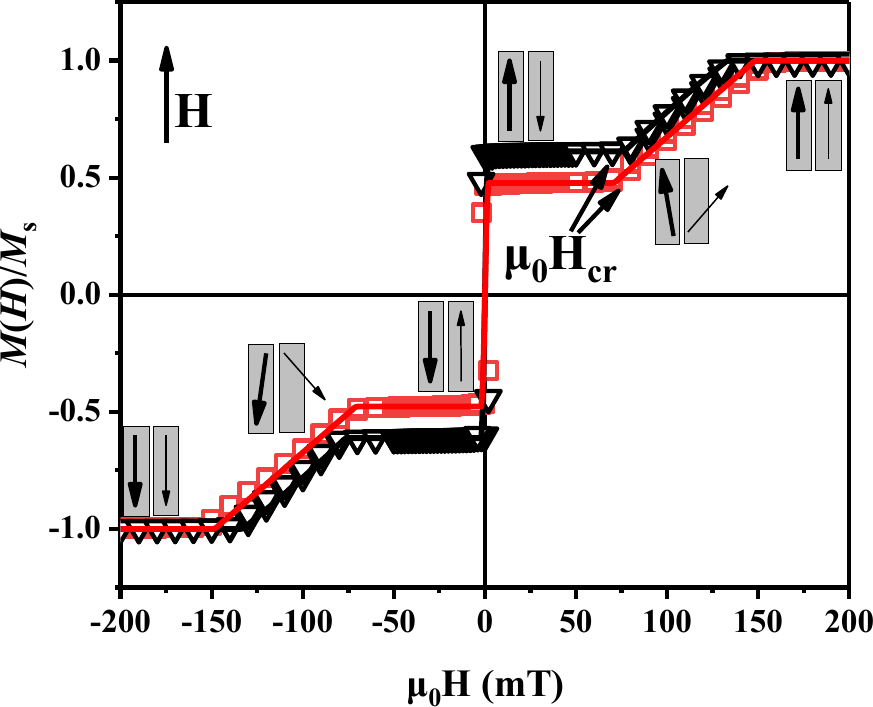}
		\vspace{10pt}
		\label{fig:SQUID}
	\end{subfigure}
	\begin{subfigure}[!t]{\linewidth}
		\captionsetup{labelfont=bf,font=large}
		\caption{}
		\includegraphics[width=\linewidth]{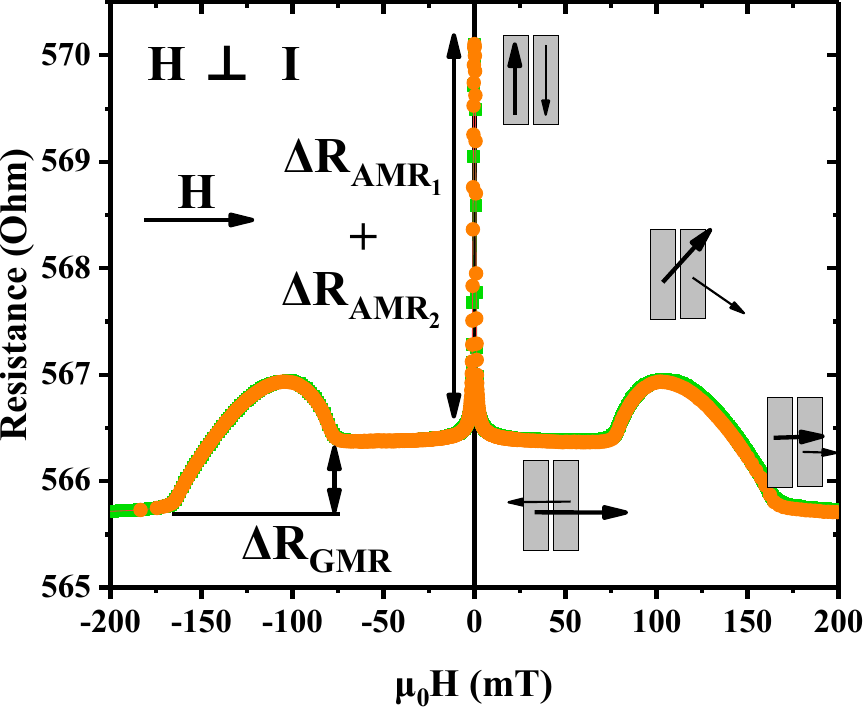}
		\label{fig:MR}
	\end{subfigure}
	\vspace{-20pt}
	\caption{\textbf{(a)} \emph{M-H} loops for Py(\SI{3}{\nano\meter})/Ru(\SI{0.85}{\nano\meter})/Py($d_\mathrm{Py}$ nm), where red squares and black triangles represent $d_\mathrm{Py}$ = \SI{6}{\nano\meter} and \SI{9}{\nano\meter}, respectively. The field is applied along the induced magnetic easy axis. The solid lines of the corresponding colors are the fits according to Eqs. \eqref{eq:Total_energy}, \eqref{eq:Magnetometry_fitting}. The black arrows schematically indicate the directions of the magnetic moment of each layer (the thicker arrow corresponds to the layer with the larger thickness). \textbf{(b)} Magnetoresistance loop for Py(\SI{3}{\nano\meter})/Ru/Py(\SI{6}{\nano\meter}), where green and orange colors denote the two different magnetic field sweep directions.}
	\label{fig:SQUID_MR}
	\vspace{10pt}
\end{figure}

The samples were fabricated using DC magnetron sputtering on 4-inch SiO$_{2}$ substrates in a ``Shamrock'' sputter deposition system. 
A small magnetic field was applied during deposition to induce a magnetic easy axis. 
The base pressure in the chamber was below $3\times 10^{-7}$ mbar. 
The full stack structure is Si/SiO$_2$/Ta(\SI{5}{\nano\meter})/Py(\SI{3}{\nano\meter})/Ru(\SI{0.85}{\nano\meter})/Py($d_\mathrm{Py}$) /Ru(\SI{3}{\nano\meter}). The Ru thickness of \SI{0.85}{\nano\meter} provides the strongest AF coupling while maintaining a continuous layer growth. 
The top-most Py layer thicknesses $d_\mathrm{Py}$ were \SI{6}{\nano\meter} and \SI{9}{\nano\meter}. For electrical measurements, chips were patterned into Hall-bar shapes with in-plane dimensions of \SI{300}{\micro\meter}$\,\times\,$\SI{10}{\micro\meter} by photolithography and ion milling.
Electrical contacts to the Hall bar were fabricated of Cr(\SI{5}{\nano\meter})/Au(\SI{125}{\nano\meter}), using UV lithography and lift-off.
An example of normalised \emph{M-H} hysteresis loops of extended films with $d_\mathrm{Py}$ = \SI{6}{\nano\meter} and \SI{9}{\nano\meter} are shown in Fig. \ref{fig:SQUID}.

Red squares and black triangles represent $d_\mathrm{Py}$ = \SI{6}{\nano\meter} and \SI{9}{\nano\meter}, respectively. 
The magnetsation direction of both layers are indicated by the black arrows. 
Statically, the \emph{M-H} response can be divided into distinct regimes based on the magnetization directions of both layers.
In the \emph{saturation regime} (above \SI{130}{\milli\tesla}), the magnetic moments are aligned with each other and saturated along the external field direction.
This occurs when the applied magnetic field has overcome both the anisotropy and the IEC.
As the field is reduced, within the range of \SI{130}{\milli\tesla} to \SI{76}{\milli\tesla}, the magnetization directions of the two magnetic layers enter the \emph{spin-flop regime} \cite{Belmequenai_2008, Py_Ru_Py_SAF_Dyn_notExpl}.
The magnetic moment of the thicker layer tilts slightly from the equilibrium position, whereas that of the \SI{3}{\nano\meter} layer gradually rotates in the plane of the sample toward antiparallel orientation relative to the \SI{9}{\nano\meter} layer and the applied field. 
Finally, at low magnetic fields (below \SI{76}{\milli\tesla}), there is an overall reduced net magnetic moment due to the colinear but antiparallel alignment of both layers, and the structure is within the \emph{antiferromagnetically (AF) coupled regime}.
The critical field between the AF-coupled and spin-flop regimes is labelled as $\mu_0H_\mathrm{cr}$ in the figure.

\begin{table*}[!t]
	\caption{Summary of the results of SQUID magnetometry fitting according to Eqs. \eqref{eq:Total_energy}, \eqref{eq:Magnetometry_fitting}. The sample structure is given in the left-most column, where ``.." represents the substrate and buffer layer, Si/SiO$_2$/Ta(5 nm). $J_1$, $J_2$ are the bilinear and biquadratic coupling constants. $M_{s_1}$, $M_{s_2}$ are the magnetizations of the 9 nm and 3 nm layers, respectively. }
	\begin{tabular}{l*{4}c}
		\hline
		\hline
		Sample & $J_1$ ($\mu$J/m\textsuperscript{2}) & $J_2$ ($\mu$J/m\textsuperscript{2}) & $M_{s_1}$ (kA/m) & $M_{s_2}$ (kA/m) \\
		\hline
		../Py(3 nm)/Ru(0.85 nm)/Py(6 nm)/Ru(3 nm) & -141$\pm 3$ & -3$\pm 3$ & 610$\pm 15$ & 432$\pm 15$\\
		&&&& \\
		
		../Py(3 nm)/Ru(0.85 nm)/Py(9 nm)/Ru(3 nm) & -131$\pm 3$ & -1$\pm 3$ & 577$\pm 15$ & 420$\pm 15$\\
		\hline
		\hline
	\end{tabular}
	\label{table:SQUID}
	\vspace*{20pt}
\end{table*}

The data is modelled in the following way.
The interlayer exchange energy per unit area is \cite{Rezende98,Belmequenai_2007}:
\begin{equation}
	E_{\rm IEC} = -J_{1}\frac{\mathbf{M}_{1}\cdot \mathbf{M}_{2}}{M_{s_1} M_{s_2}}               
	-J_2\left(  \frac{\mathbf{M}_{1}\cdot \mathbf{M}_{2}}{{M_{s_1} M_{s_2}}}   \right)^2
	\label{eq:IEC}
\end{equation}
where $J_1$, $J_2$ are, respectively, the bilinear and biquadratic coupling constants.
$\mathbf{M}_{1}$, $\mathbf{M}_{2}$ are the magnetizations of the two Py layers, with $M_{s_1}$, $M_{s_2}$ corresponding to the saturation values.
From the arguments of total energy minimization, the negative sign of $J_1$ favors anti-parallel (AF) coupling, while a negative $J_2$ leads to a \SI{90}{\degree} equilibrium state.
If $J_{1,2} > 0$, a parallel alignment is favored.

In our geometry, both magnetizations are confined to the plane of the films. 
Therefore, the magnetic free energy of the system per unit area is:
\begin{multline}
	E_T = \\ 
	\sum_{i=1,2} d_i\left[\left( K_{u_{i}}\sin^2(\alpha_i)- \mu_0 M_{s_i} H \cos(\alpha_{0}-\alpha_i) \right) \right]- \\
	-J_1 \cos(\alpha_1-\alpha_2) - J_2 \cos^2(\alpha_1-\alpha_2).
	\label{eq:Total_energy}
\end{multline}
In Eq. (\ref{eq:Total_energy}), $d_i$ are the thicknesses of the layers, $K_{u_{i}}$ their uniaxial magnetocrystalline anisotropy constants, $\mu_0 H$ is the amplitude of the external magnetic field applied at angle $\alpha_0$, and $\alpha_i$ are the angles between magnetizations and the uniaxial anisotropy. 
Minimization of Eq.~\eqref{eq:Total_energy} with respect to $\alpha_i$ yields the equilibrium directions of both magnetizations ($\alpha_{1_{e}},\alpha_{2_{e}}$) for all applied field values. The magnetization for each field value is then calculated as~\cite{Belmequenai_2007}:
\begin{multline}
\frac{M(\textit{H})}{M_{s}} = \frac{d_1 M_{s_1}\cos[\alpha_0-\alpha_{1_{e}}\left(J_1,J_2\right) ]}{d_1 M_{s_1}+d_2 M_{s_2}} +\\
+\frac{d_2 M_{s_2}\cos[\alpha_0-\alpha_{2_{e}}\left(J_1,J_2\right)]}
{d_1  M_{s_1}+d_2 M_{s_2}},
\label{eq:Magnetometry_fitting}
\end{multline}
where $M_{s}$ is the total saturation magnetization of the stack. 
The bilinear and biquadratic coupling constants (\textit{J}$_{1}$, \textit{J}$_{2}$) are determined by fitting magnetometry loops  according to Eqs.~\eqref{eq:Total_energy} and \eqref{eq:Magnetometry_fitting}.
The solid lines in Fig. \ref{fig:SQUID} are fits according to those equations.

Fig. \ref{fig:MR} shows the magnetoresistive response for a patterned Hall bar structure with $d_\mathrm{Py}$ = \SI{6}{\nano\meter}. 
Here, the magnetic field is applied in the plane of the bar, perpendicular to its long axis and the current direction.
The green and orange lines represent the field sweep directions and show that as in the \emph{M-H} loops, there is no large hysteretic behavior in the spin-flop regime.
Again, the black arrows indicate the magnetization directions of the two Py layers. 
The saturated, spin-flop, and AF-coupled regimes occur at the same field values as in Fig. \ref{fig:SQUID}.
In the absence of any applied field, the magnetizations lie along the bar and parallel to the current direction, leading to a high resistance due to the AMR contributions, $\Delta R_\mathrm{AMR_{1}}$ and $\Delta R_\mathrm{AMR_{2}}$, of the two Py layers. 
Upon the application of a weak magnetic field above the shape anisotropy field of the structure, $\mu_0H > \SI{5}{\milli\tesla}$, the magnetizations are aligned perpendicular to the current direction, and thus, the resistance is lowered.
At high fields, in the saturated regime, the further drop in resistance is due to the giant magnetoresistance (GMR) effect. 
This is minimized when the magnetizations are aligned parallel.
The increase and subsequent drop in the resistance within the spin-flop regime is due to the competition of $\Delta R_\mathrm{AMR_{1}}$, $\Delta R_\mathrm{AMR_{2}}$ and the GMR effect.
 
\subsection{Magnetization dynamics and detection}


The dynamical behavior of the magnetizations is calculated using the system of coupled Landau-Lifshitz (LL) equations of motion. The equation for the $i$-th layer can be written as:
\begin{equation}
\mathbf{\dot M}_i = -\gamma\mu_0\left[\mathbf{M}_i \times \mathbf{H}^{\rm eff}_i\right].
\label{eq:LLG}
\end{equation}
Here, $\mathbf{H}^{\rm eff}_i$ is the effective magnetic field given by:
\begin{equation}
\mathbf{H}^{\rm eff}_i =\mathbf{H}+ \mathbf{H}_{u_i} + \mathbf{H}_{d_i}+ \mathbf{H}_{s} + \mathbf{H}_{\mathrm{IEC}_i} (J_1,J_2),
\label{eq:EFF_fields}
\end{equation}
and consisting of the externally applied field, $\mathbf{H}$, the uniaxial anisotropy field, $\mathbf{H}_{u_i}$, the demagnetizing field, $\mathbf{H}_{d_i}$, the surface anisotropy field $ \mathbf{H}_{s}$ and IEC field $\mathbf{H}_{\mathrm{IEC}_i}$. 
In the linear approximation, the magnetization can be written as $\mathbf{M}_i=\mathbf{m}_i+M_{s_i} \hat{\kappa}_{e}$, where
$\hat{\kappa}_{e}$ is the equilibrium direction of the magnetization, 
$\lvert\mathbf{m}_i\rvert << M_s$ is the time-varying component of $\mathbf{M}$, perpendicular to $\hat{\kappa}_e$.
Using this approach, the effective fields on both layers can be analytically calculated. 
An explicit expression for the different contributions of the effective fields can be found in Ref. \cite{Rezende98} for a finite value of the wave vector. 
In our case, we are interested in the limit of the uniform precession ($\mathbf{k} = 0$).

In order to explain the non-monotonic $f(\mathbf{H})$ response, we will analyze the dynamic energies per unit area ($\epsilon$) associated to the bilinear IEC and Zeeman terms. These energies are:
\begin{multline}
\epsilon_{\rm IEC} = \sum_{i=1,2}\frac{J_1 \cos(\alpha_{1_{e}}-\alpha_{2_{e}})}{2}(m_{{\rm IP}_i}^2+m_{{\rm OOP}_i}^2)-
\\-J_1 \cos(\alpha_{1_{e}}-\alpha_{2_{e}})(m_{{\rm IP}_1}m_{{\rm IP}_2}+m_{{\rm OOP}_1}m_{{\rm OOP}_2})
\label{eq:IEC2}
\end{multline}

and 
\begin{multline}
\epsilon_{\rm Z} = \sum_{i=1,2}\frac{\mu_0 H M_{s_i} d_i \cos(\alpha_{0}-\alpha_{i_{e}})}{2}(m_{{\rm IP}_i}^2+m_{{\rm OOP}_i}^2),
\label{eq:Z2}
\end{multline}
where $m_{{\rm IP}_i}$, $m_{{\rm OOP}_i}$ are the in-plane (IP) and out-of-plane (OOP) dimensionless components of the dynamical magnetization component $\mathbf{m}_i$.
Although all energetic contributions influence the dynamics of the system, we demonstrate that Eqs. (\ref{eq:IEC2}) and (\ref{eq:Z2}) mainly describe $f(\mathbf{H})$.

\begin{figure}[!t]
	\centering
	\begin{subfigure}[!t]{0.5\textwidth}
		\centering
		\captionsetup{labelfont=bf,font=large}
		\caption{}
		\includegraphics[width=0.8\linewidth]{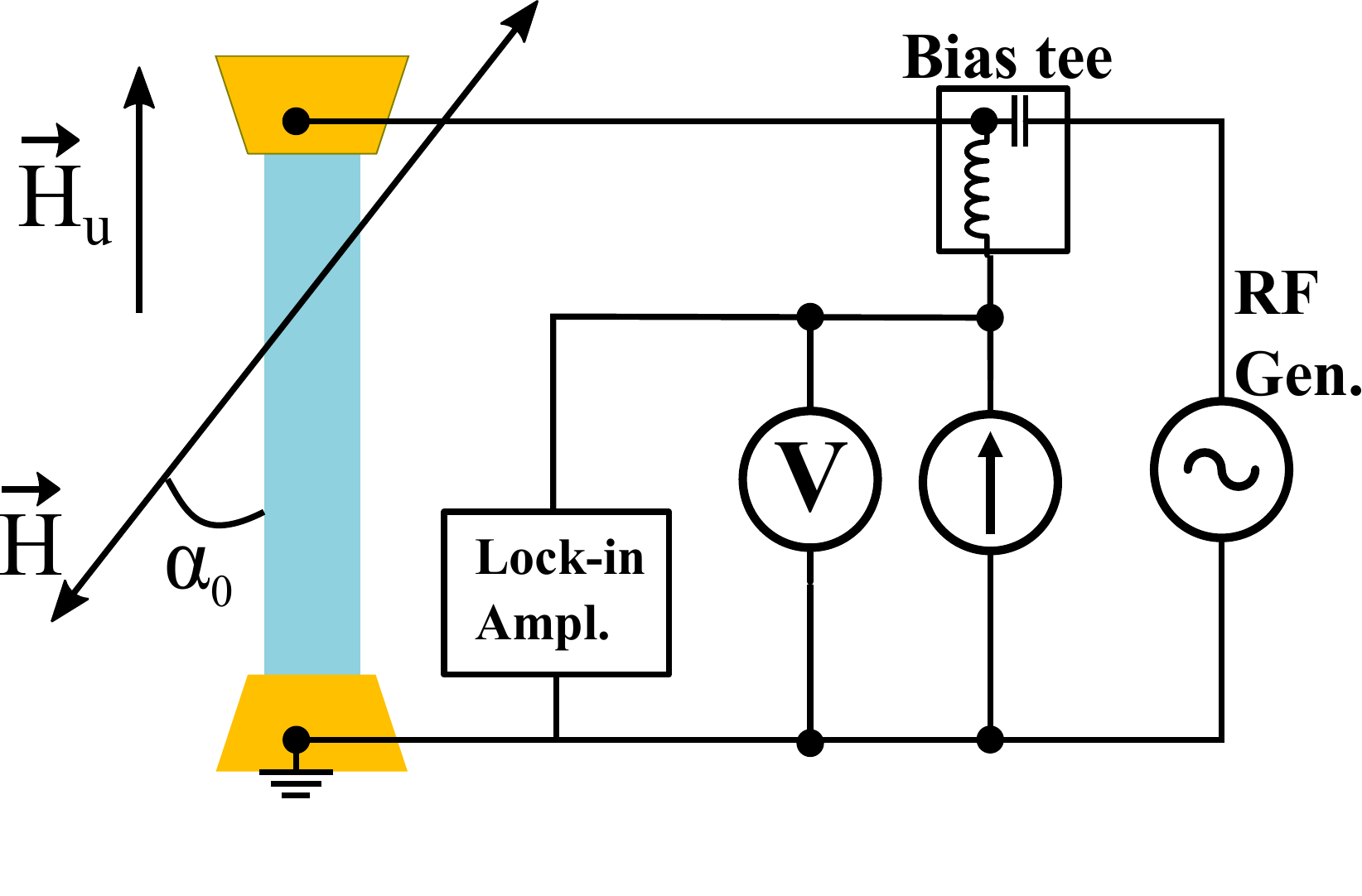}
		\label{subfig:ED_FMR_setup}
		\vspace{20pt}
	\end{subfigure}
	\begin{subfigure}[!t]{0.5\textwidth}
		\centering
		\captionsetup{labelfont=bf,font=large}
		\caption{}
		\includegraphics[width=0.8\linewidth]{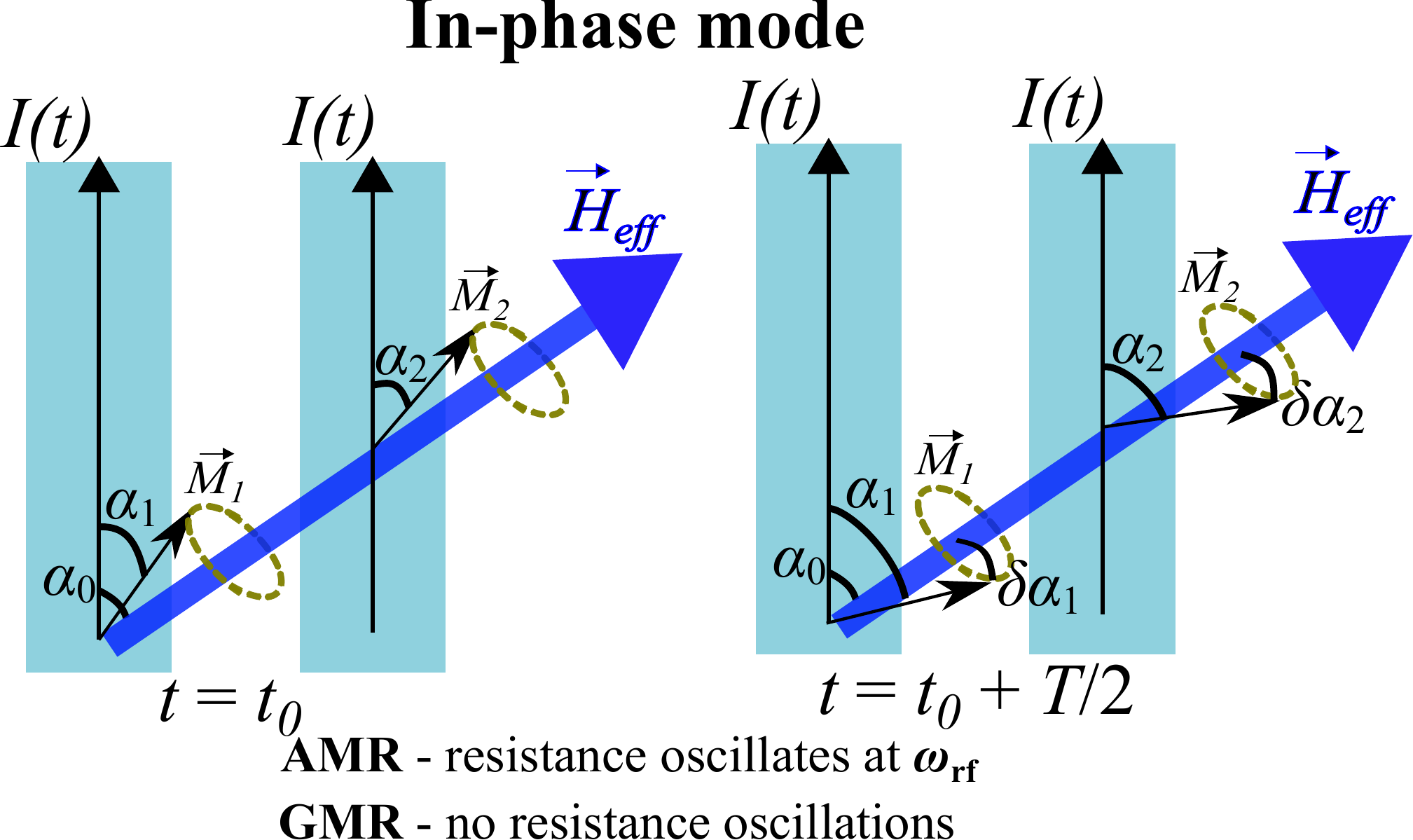}
		\label{subfig:MR_EFFECTS_IPhmode}
		\vspace{20pt}
	\end{subfigure}
	\begin{subfigure}[!t]{0.5\textwidth}
	\centering
	\captionsetup{labelfont=bf,font=large}
	\caption{}
	\includegraphics[width=0.8\linewidth]{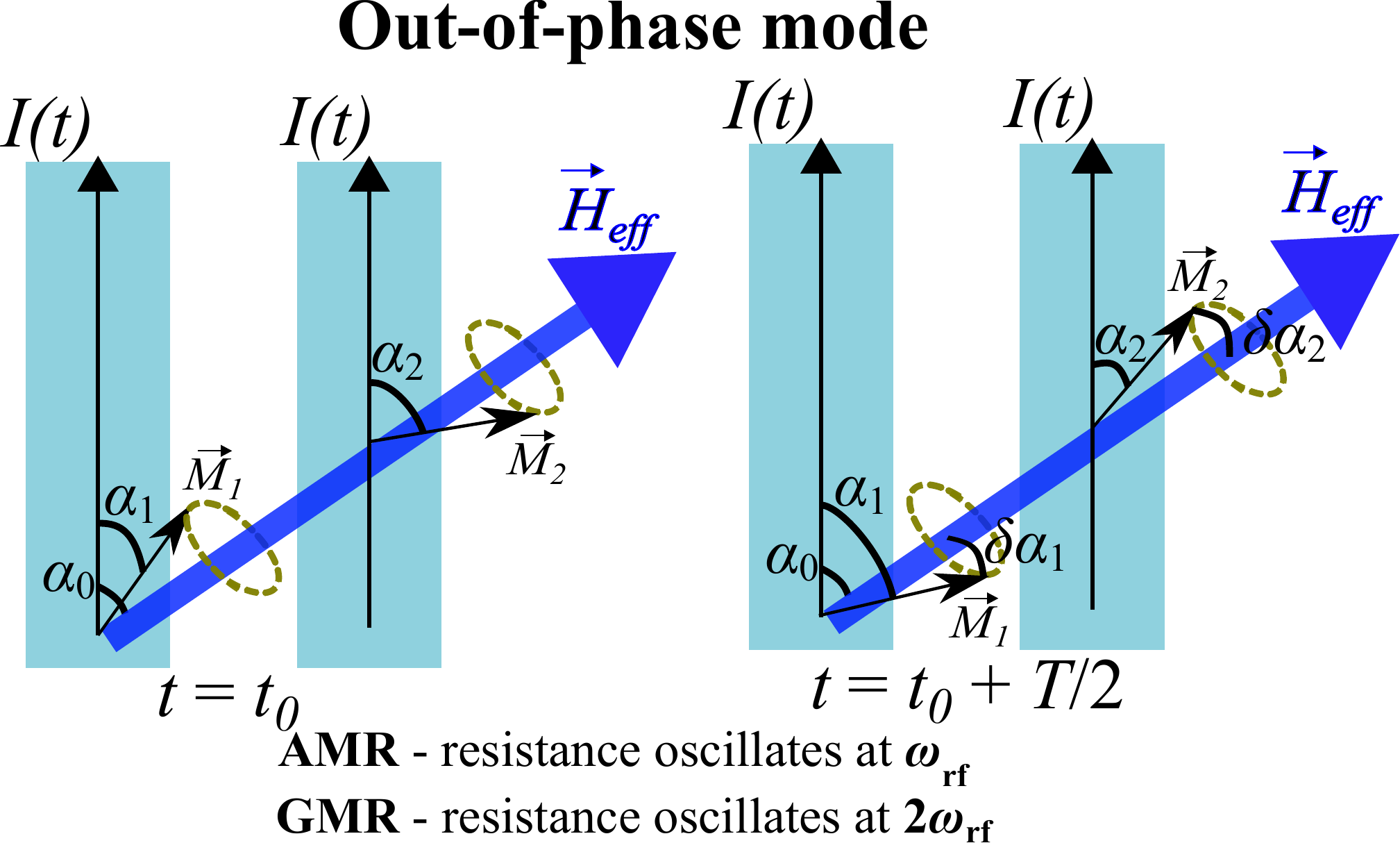}
	\label{subfig:MR_EFFECTS_OOPhmode}
	\end{subfigure}

	\caption{\textbf{(a)} Representation of the ED-FMR setup used. The field was applied at 45$^{\circ}$ to the strip (current) direction.
	\textbf{(b), (c)} two relevant magnetoresistance effects, which can be used. Only AMR gives a contribution at the excitation frequency, which is necessary to rectify the signal.}
	\vspace{20pt}
	\label{fig:setup}
\end{figure}

Fig. \ref{fig:setup} shows the experimental set up.
The patterned sample is depicted as a light blue bar.
The external magnetic field $H$ is applied in the sample plane, at an angle $\alpha_0$ to the current direction.

Frequency-swept spectra at fixed magnetic field are recorded. The external magnetic field is modulated by an additional set of coils installed on the magnet pole-shoes. Thus, the field-derivative of the rectified signal is measured. 
This approach was shown to perform much better compared to amplitude-modulated frequency-swept ED-FMR \cite{Barsukov_fieldmod}, and allows us to detect both modes for the same static magnetic configuration. 
Such an investigation is not possible when sweeping the magnetic field.

Both AMR and GMR are changing during the resonance and contribute to the detected signals. However, GMR varies as a cosine of the angle between the two magnetizations. Hence, for in-phase modes, no change in resistance is expected. For out-of-phase mode, the frequency of resistance oscillations due to GMR is $2 \omega_\mathrm{RF}$ and no rectification to DC occurs.
Note that $\omega_\mathrm{RF}$ is the excitation frequency and not the modulation frequency provided by the lock-in.
%
The total rectified voltage in the case of the saturated layers,  is given by \cite{EDFMR_overview}:
 \begin{multline}
 \langle U_\mathrm{DC}\rangle\propto I_\mathrm{RF}^2[\Delta R_\mathrm{AMR_{1}}\sin(2\alpha_{1}) \delta\alpha_{1}\pm \\
 \pm\Delta R_\mathrm{AMR_{2}}\sin(2\alpha_{2}) \delta\alpha_{2}],
 \label{eq:AMR_voltage}
 \end{multline}
 where $I_\mathrm{RF}$ is the amplitude of the RF current induced by the RF generator, $\Delta R_\mathrm{AMR_{1}}$, $\Delta R_\mathrm{AMR_{2}}$ are the amplitudes of the total resistance changes due to the AMR effect and $\delta\alpha_{1}$, $\delta\alpha_{2}$ are the amplitudes of the small changes in the angle between the magnetization and the current direction for each layer, when the resonance condition if fulfilled. ``+" and ``-" signs correspond to the in-phase and out-of-phase precession cases, respectively.
 The changes in the equilibrium angles are directly connected to the amplitude of the above mentioned in-plane dynamical magnetization components as:
 \begin{equation}
 	\delta\alpha_i = \frac{m_\mathrm{IP_i}}{M_{s_i}}.
 	\label{eq:angletoMip}
 \end{equation}
More details about spin-rectification effects can be found in Refs. \cite{EDFMR_overview, MeckingEDFMR}.

\subsection{FMR response and the role of the dynamical energy}

\begin{figure}[!h]
	
	\begin{subfigure}{0.86\linewidth}
		\centering
		\captionsetup{labelfont=bf,font=large}
		\caption{}
		\includegraphics[width=\linewidth]{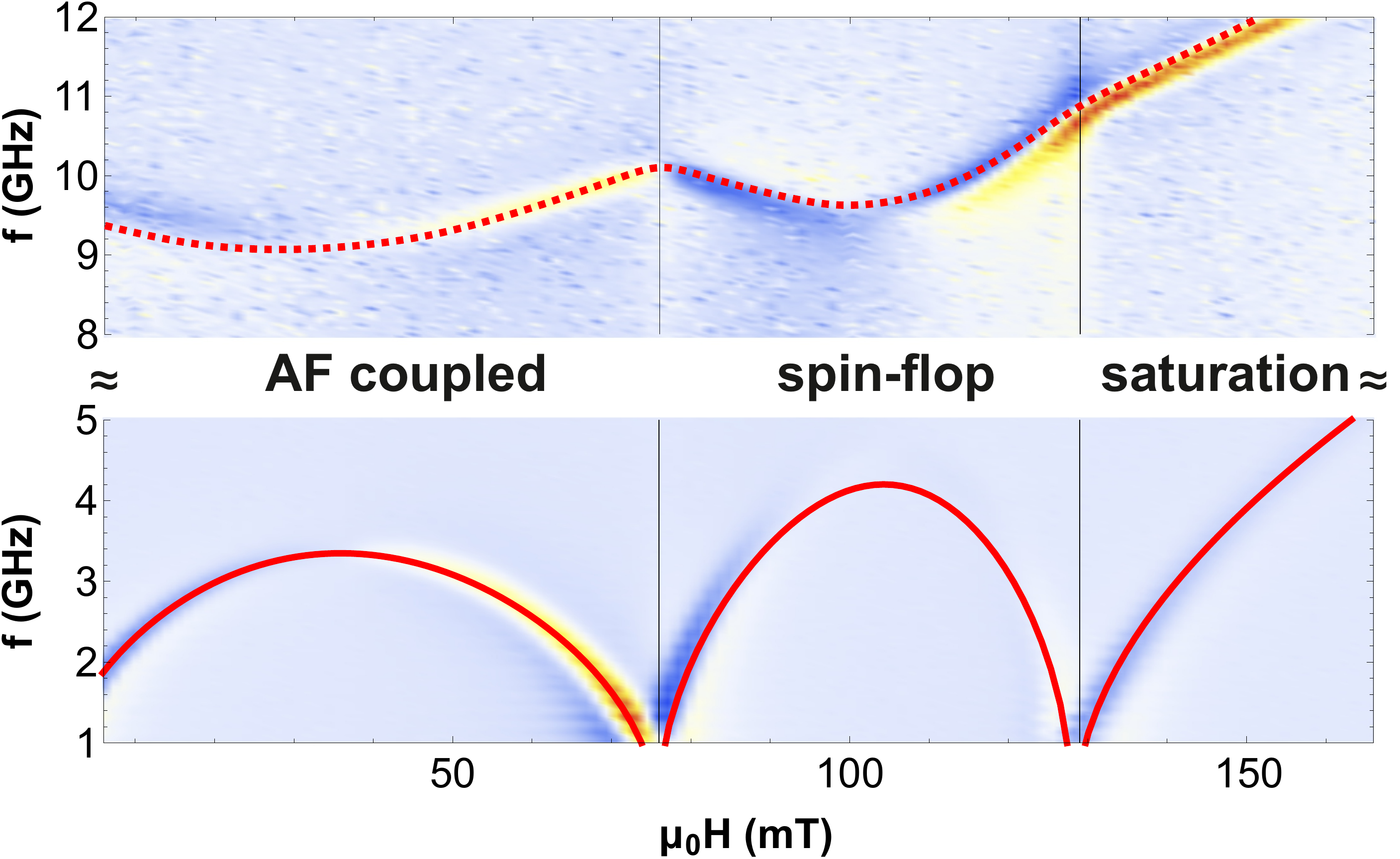}
		\vspace{10pt}
		\label{fig:9_3EDFMR}
	\end{subfigure}
	\begin{subfigure}{0.86\linewidth}
				\captionsetup{labelfont=bf,font=large}
				\caption{}
				\includegraphics[width=\linewidth]{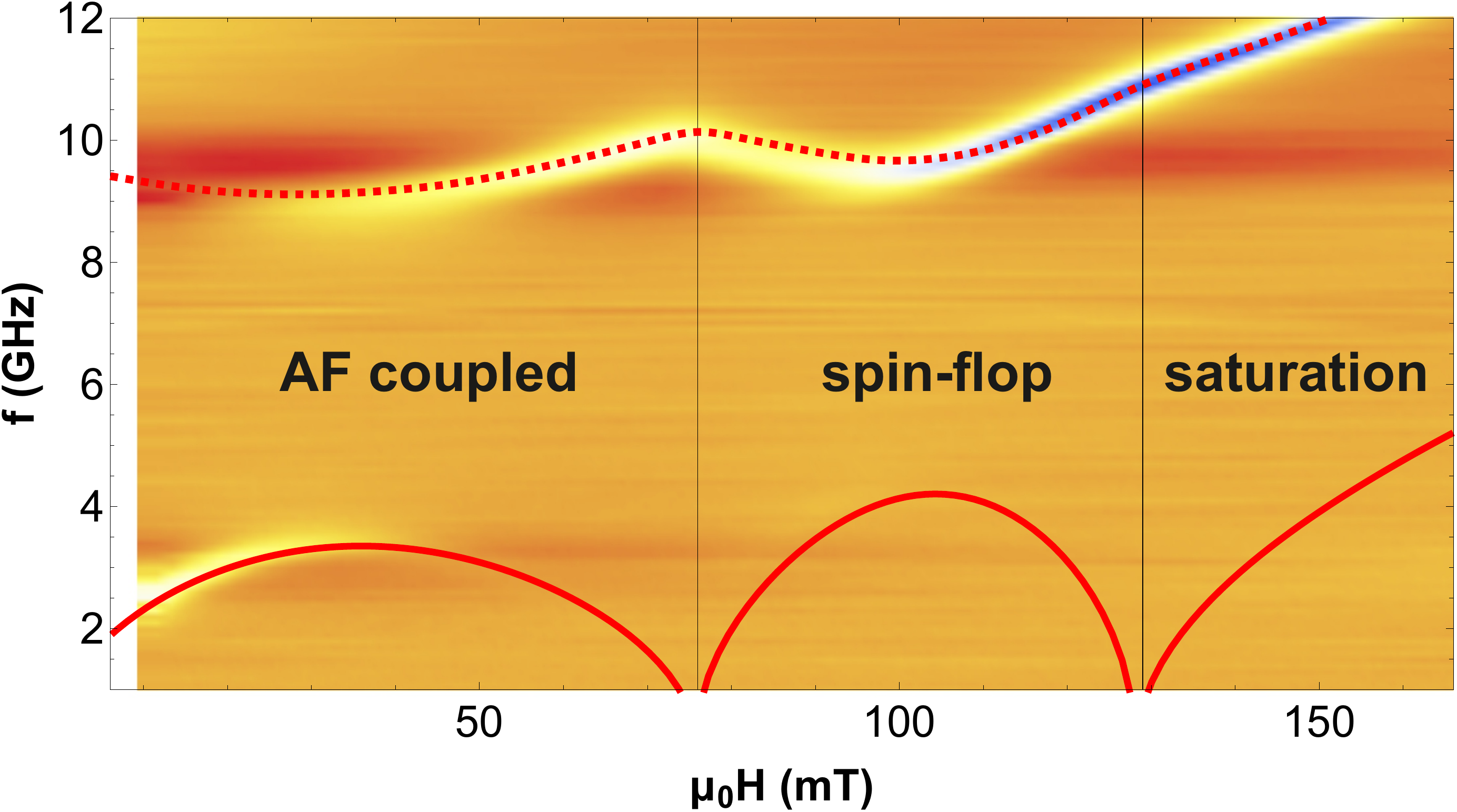}
				\label{fig:9_3FMR}
	\end{subfigure}
	\begin{subfigure}{0.86\linewidth}
				\captionsetup{labelfont=bf,font=large}
				\caption{}
				\includegraphics[width=\linewidth]{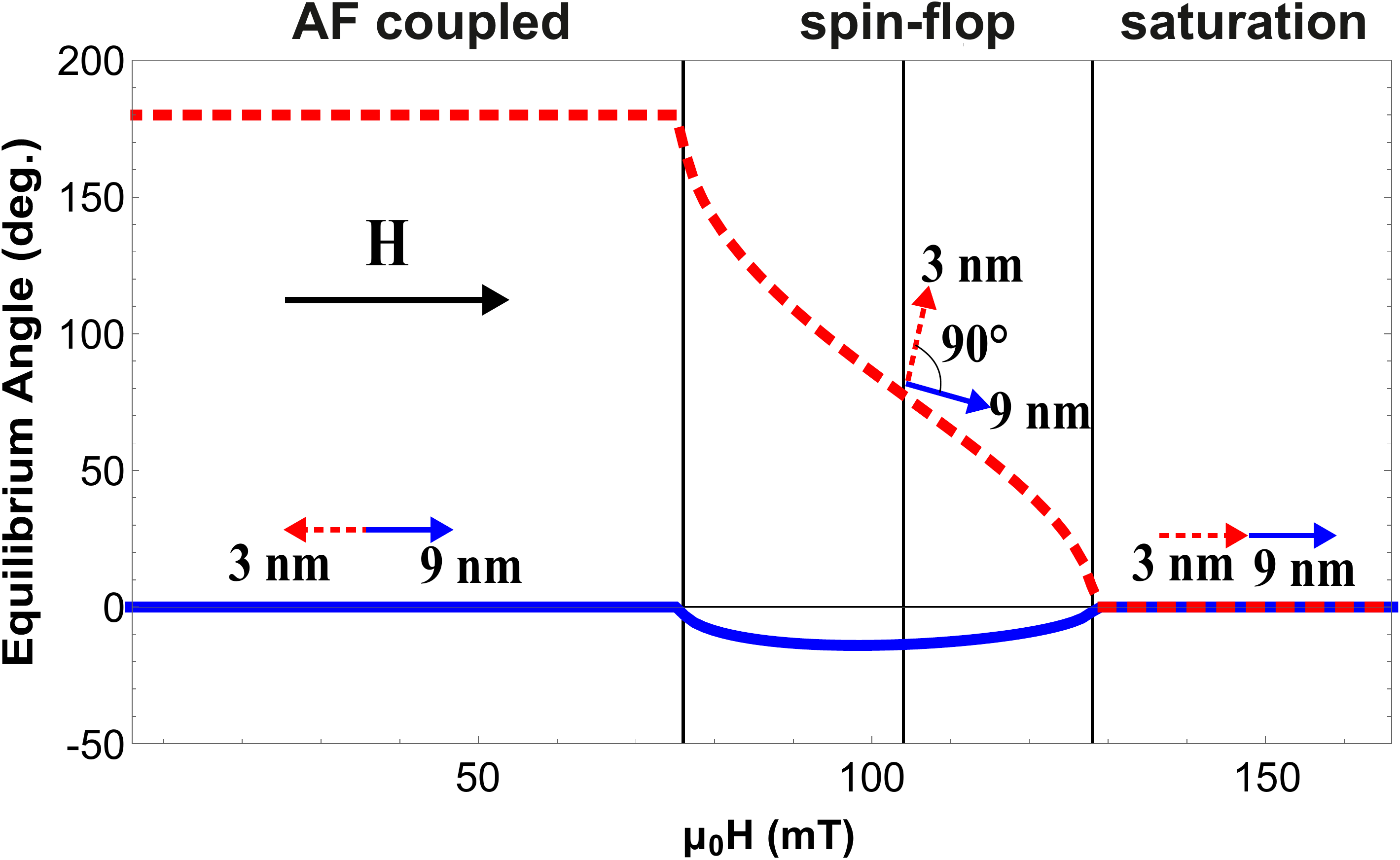}
				\label{fig:Eq_angles}
				
	\end{subfigure}
	\vspace{-15pt}
	\caption{ED-FMR \textbf{(a)}  and VNA-FMR \textbf{(b)}   measurements for Py(3 nm)/Ru/Py(9 nm). Solid and dashed lines represent calculations according to the Eqs. \eqref{eq:LLG}, \eqref{eq:EFF_fields}. Parameters for the modeling ($J_i$, $M_i$) were taken from SQUID-VSM data analysis (Table \ref{table:SQUID}), except in the case of the 9/3 sample, $J_1 $ was decreased to \SI{-128}{\micro\joule/\square\meter} to achieve better agreement with the dynamic data. Colors represent relative amplitudes of the resonances. \textbf{(c)} The dependence of the equilibrium angles of the magnetizations of the two Py layers versus applied field. Solid blue lines correspond to the thicker \SI{9}{\nano\meter} layer, and dashed red to the \SI{3}{\nano\meter} layer.} 
	\label{fig:Dynamics}
	\vspace{20pt}
\end{figure}

We now turn to the results of both ED-FMR and VNA-FMR, shown in Figs.~\ref{fig:9_3EDFMR} and \ref{fig:9_3FMR}, respectively. 
Two modes are observed, as expected. 
Our model based on the coupled LL equations [Eqs.~\eqref{eq:LLG},~\eqref{eq:EFF_fields}], provides a good fit for both modes [dashed and solid lines in Figs.~\ref{fig:9_3EDFMR},~\ref{fig:9_3FMR}].
Three regions that correspond to the AF-coupled, spin-flop and saturated regimes for both the high- and low-frequency modes can be clearly identified.
We also plot the equilibrium angles of magnetizations of both Py layers ($\alpha_{1_{e}}$, $\alpha_{2_{e}}$), calculated from Eqs. (\ref{eq:Total_energy}, \ref{eq:Magnetometry_fitting}) in Fig. \ref{fig:Eq_angles}, using the parameters determined above (see Table \ref{table:SQUID}). 
A critical point, $\mu_0 H = \SI{104}{\milli\tesla}$, is found to correspond to a crossover of both modes occurring midway through the spin-flop field where the two magnetizations are at \SI{90}{\degree} to each other.

 \begin{figure*}[t]
	\begin{subfigure}[b]{0.5\textwidth}
		\captionsetup{labelfont=bf,font=large}
		\caption{}
		\includegraphics[width=\linewidth]{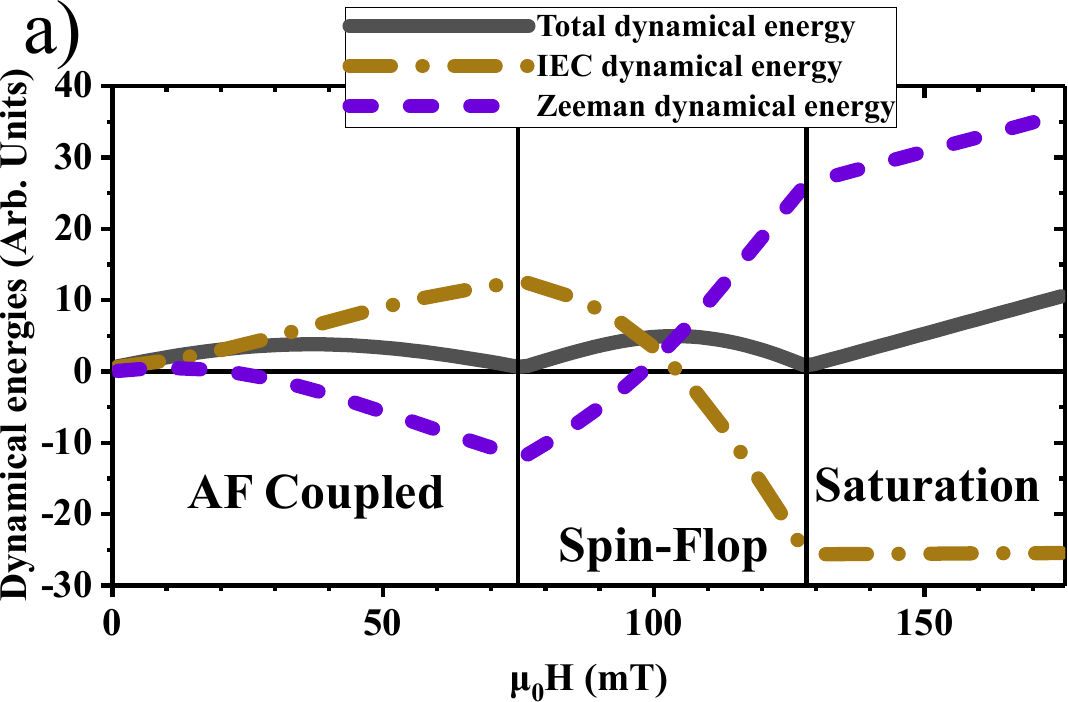}
		\label{subfig:LF_dyn_energies}
	\end{subfigure}
	\begin{subfigure}[b]{0.5\textwidth}
		\captionsetup{labelfont=bf,font=large}
		\caption{}
		\includegraphics[width=\linewidth]{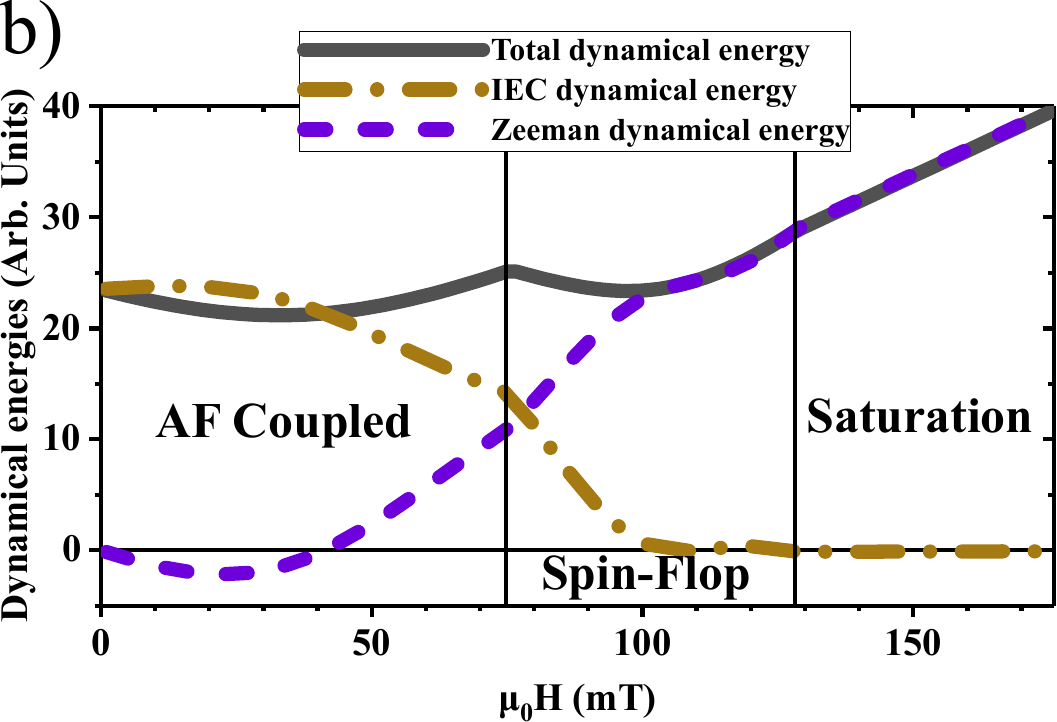}
		\label{subfig:HF_dyn_energies}
	\end{subfigure}
	\caption{Dynamical part of the total energy for the Py(\SI{3}{\nano\meter})/Ru/Py(\SI{9}{\nano\meter}) sample in case of low-frequency mode (a) and high-frequency mode (b). Dash-dotted brown curves correspond to the dynamical IEC energy, Eq. (\ref{eq:IEC2}), and dashed purple curves correspond to the dynamical Zeeman energy, Eq. (\ref{eq:Z2}). Solid black curves represent the total dynamical energy, which is the sum of the two. }
	\vspace{30pt}
	\label{fig:dynamical_en}
\end{figure*}

As shown in Refs.~\cite{Belmequenai_2007} and~\cite{Belmequenai_2008}, the precession phase difference of the magnetizations of the two layers does not remain constant throughout the whole field range. At low fields, in the AF-coupled regime and midway into the spin-flop regime, the relative angle between the magnetizations is above \SI{90}{\degree} [see Fig. \ref{fig:Eq_angles}], and thus the excitation of the in-phase precession does not involve the IEC energy to a high degree. This mode has a lower frequency. 
The out-of-phase precession, up to the same point, drives  the magnetizations away from the antiparallel configuration; consequently, the system has to overcome a certain amount of the IEC energy, and thus, the mode exhibits a larger frequency.

Above this point, i. e.,  when the relative angle between magnetizations is below \SI{90}{\degree}, the opposite is observed. In particular, the in-phase precession exhibits a higher frequency, than the out-of-phase precession. This happens due to the fact that now the in-phase precession keeps the layers closer to, or in, the parallel state.
This is energetically unfavorable from the point of view of IEC and thus in-phase precession now corresponds to the higher frequency. 
Due to this interchange of the mode character, we will no longer refer to the modes as ``optic'' or ``acoustic'' modes to avoid confusion. 
From now on, the modes will be referred to simply as ``low-frequency'' and ``high-frequency'' modes, with the phase difference directly specified when required. 


Although one can explain the transition in the precession state in the spin-flop regime from the static energy considerations only, it is not possible to explain the drastic changes in the $f(\mathbf{H})$ dependence. 
The presence of the two pronounced maxima and minima in the AF-coupled regime, where statically nothing changes according to both, magnetometry and magnetoresistance data, is particularly striking. 
To understand this behavior, we plot the dynamic Zeeman (purple dashed line) and IEC (brown dash-dotted line) energies, as well as their sum (solid black line) versus the applied magnetic field for the low-frequency and high-frequency modes (Fig. \ref{fig:dynamical_en}) calculated according to Eqs. (\ref{eq:IEC2}) and (\ref{eq:Z2}). 
We emphasize that the plotted energies contain no static parts, they represent the differences in the total energy, on- and off-resonance.
The total dynamical energy follows the traces of the $f(\mathbf{H})$ response.
  
  \begin{figure*}[t]
  	  	\begin{subfigure}[b]{0.5\textwidth}
  		\captionsetup{labelfont=bf,font=large}
  		\caption{}
  		\includegraphics[width=\linewidth]{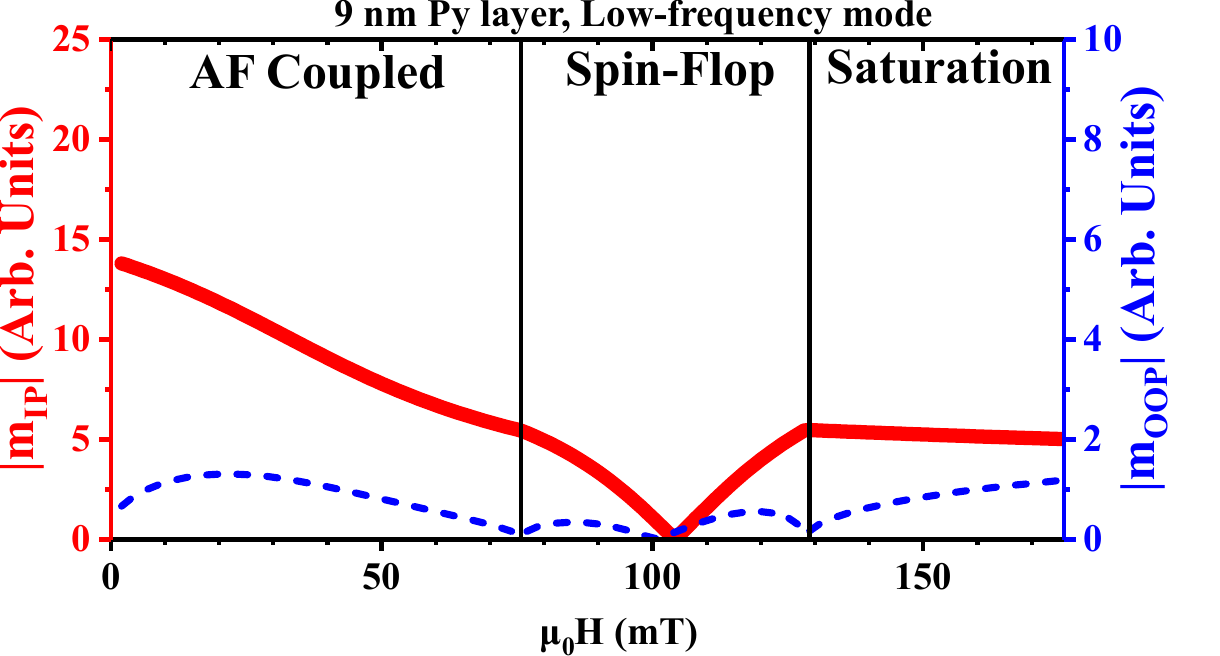}
  		\label{subfig:Traj_9_LF}
  	\end{subfigure}
  	\begin{subfigure}[b]{0.5\textwidth}
  		\captionsetup{labelfont=bf,font=large}
  		\caption{}
  		\includegraphics[width=\linewidth]{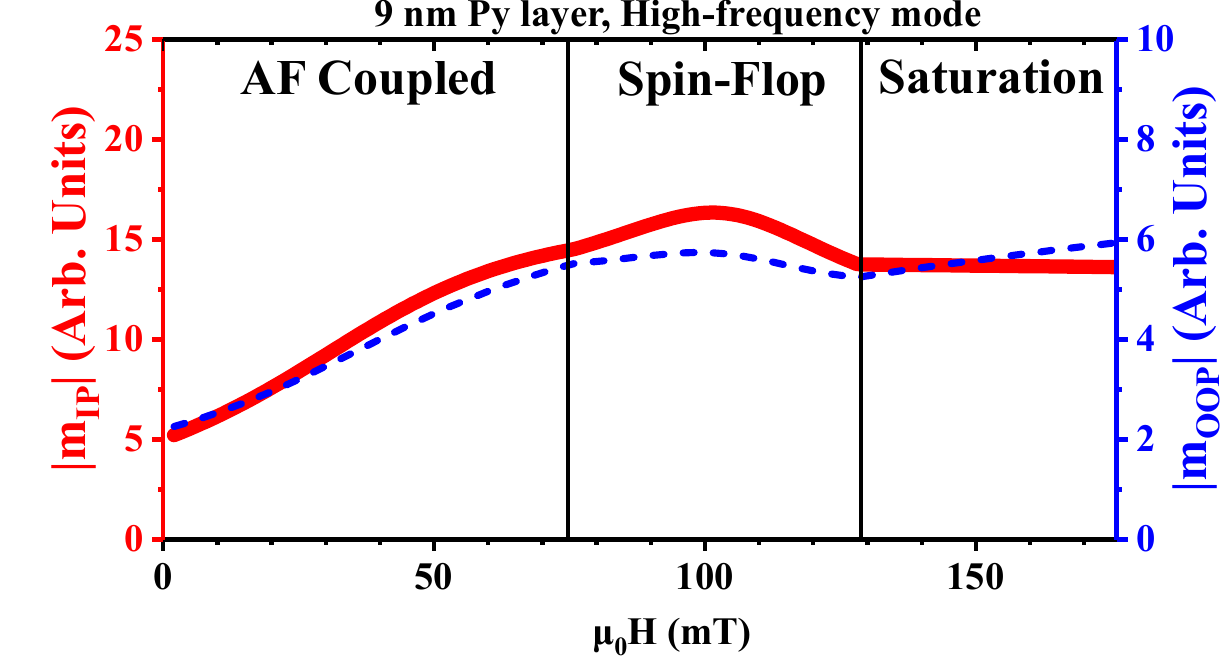}
  		\label{subfig:Traj_9_HF}
  	\end{subfigure}
  	\begin{subfigure}[b]{0.5\textwidth}
  		\captionsetup{labelfont=bf,font=large}
  		\caption{}
  		\includegraphics[width=\linewidth]{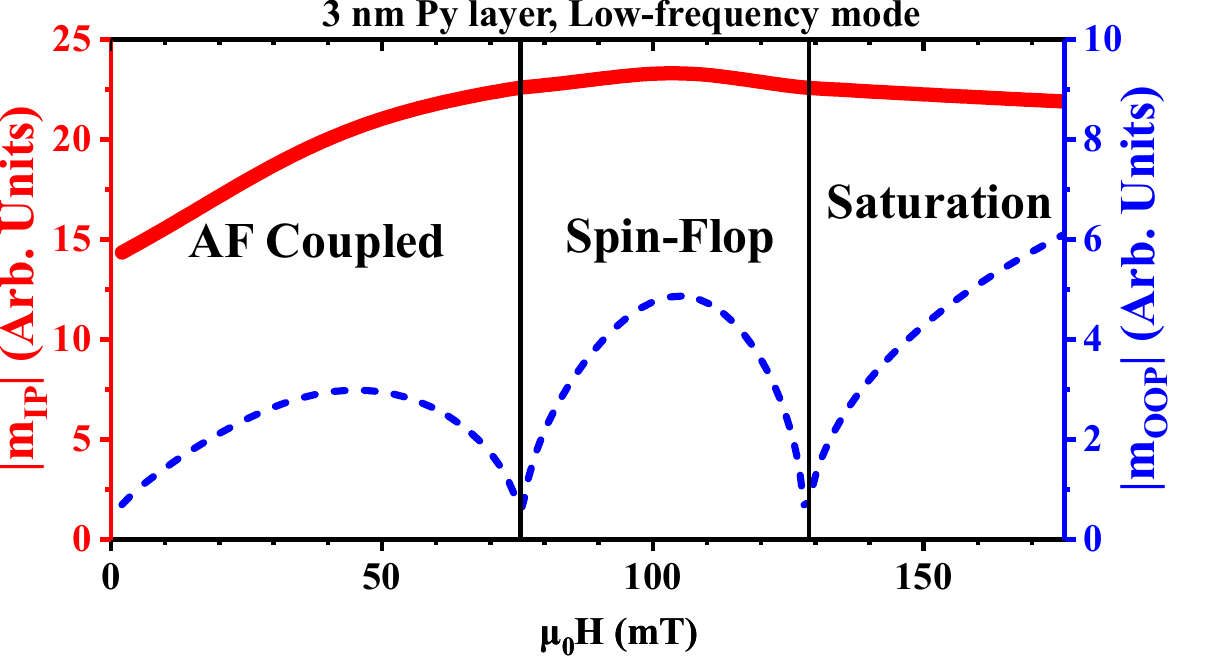}
  		\label{subfig:Traj_3_LF}
  	\end{subfigure}
  	\begin{subfigure}[b]{0.5\textwidth}
  		\captionsetup{labelfont=bf,font=large}
  		\caption{}
  		\includegraphics[width=\linewidth]{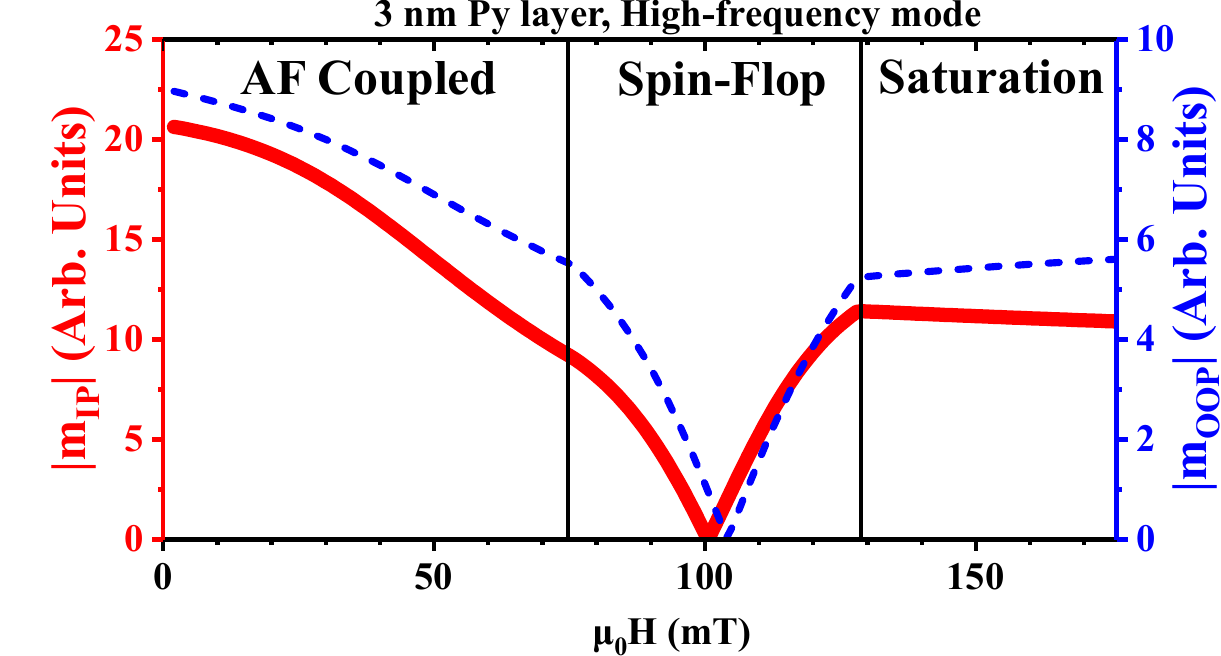}
  		\label{subfig:Traj_3_HF}
  	\end{subfigure}
  	\caption{Relative magnitudes of the in-plane (in red) and out-of-plane (in blue) dynamical components of the magnetizations of the 9 nm layer\textbf{(a)}, \textbf{(b)} and the 3 nm layer \textbf{(c)}, \textbf{(d)} for the low-frequency mode \textbf{(a)}, \textbf{(c)} and  the high-frequency mode \textbf{(b)}, \textbf{(d)}, respectively. }
  	\vspace{30pt}
  	\label{fig:Trajectories}
  \end{figure*}

We first focus on the low-frequency mode in the AF-coupled regime [Fig. \ref{subfig:LF_dyn_energies}]. As the magnetic field increases, the Zeeman energy change is different for the two layers due to the asymmetry in the total magnetic moment. The dynamical IEC energy is steadily increasing with external field. That means that the average angle between magnetizations is changing from \SI{180}{\degree} (as assumed from static considerations) to a lower value. The higher the field (as long as we are still in the AF-coupled regime), the higher the effect. Since this dependence only deals with the dynamical energies, it means that the precession cones of the layers are changing. Now, the overall dynamical Zeeman energy is decreasing. Therefore, this is the area of the precession cone of the 9 nm layer which is decreasing, whereas that of the 3 nm is increasing. After the spin-flop regime is reached, both dynamical and statical energies are changing, but it is the interplay between dynamical energies that reproduces qualitatively the behavior of the $f(\mathbf{H})$ dependence. In saturation, the dynamical IEC energy has a constant value, while the Zeeman energy increases linearly with the field, as expected. 

In the case of the high-frequency mode [Fig. \ref{subfig:HF_dyn_energies}], the dynamical IEC energy is not zero at zero magnetic field, since, as one remembers from the static considerations, in the AF-coupled regime this mode corresponds to the out-of-phase precession. As the field increases, the dynamical IEC energy steadily decreases. Due to the same Zeeman energy asymmetry as for the low frequency mode, precession trajectories of the layers are changing with the field. But now, the dynamical Zeeman energy is increasing (after a small minimum in the low fields). That means, that is the precession trajectory of the thicker layer that is increasing, whereas that of the thinner layer is decreasing (as oppose to the low-frequency mode). Although similar to the low-frequency case, in saturation, the dynamical IEC energy stays constant; the actual value is higher. This agrees with the arguments presented in the statics-based explanation. Namely, in the saturation regime, the high-frequency mode corresponds to the in-phase precession, keeping a \SI{0}{\degree} average angle between the magnetizations. This is the highest possible energy state for IEC.

We now compare the observed amplitudes with our model for both VNA-FMR and ED-FMR, starting again from the low-frequency mode.
We recall here that while the VNA-FMR signal amplitude is proportional to the inductive response of the sample, the ED-FMR amplitude is proportional to the time-dependent dynamical IP component of the magnetization, $m_\mathrm{IP}$. From the solution of the LL equation (Eq. \ref{eq:LLG}) and experimentally determined effective fields, we infer the magnitudes of the IP and OOP dynamical magnetization components for each layer for the different modes, plotted in Fig. \ref{fig:Trajectories}. The magnitudes of the IP components (red curves) evolve in the same way as was assumed based on the dynamical energies. The decrease in the IP component of the \SI{9}{\nano\meter} layer coincides with the gradual amplitude decrease of the VNA-FMR signal in this regime [Fig.~\ref{fig:9_3FMR}] \cite{Endnote_1}, further corroborating the dynamical magnetization as the source of the $f(\mathbf{H})$ response.
VNA-FMR detects the inductive response of the sample; therefore, samples with a higher magnetic volume contribute more to the absorption. 
Thus, the decrease of the precession cone angle and the consecutive decrease of the dynamic magnetization components of the \SI{9}{\nano\meter} layer dominate the overall signal intensity. For the ED-FMR measurement, on the other hand, the signal results from the resistance changes due to the precessing magnetizations. 
Therefore, for ED-FMR, the amplitude dependence is not as dramatic as observed in Fig. \ref{fig:9_3EDFMR} for the low-frequency mode. For the high frequency mode, the VNA-FMR amplitude is gradually increasing throughout the AF-coupled regime. This behavior is consistent with the increase of the dynamical trajectory of the 9 nm layer [Figs. \ref{subfig:Traj_9_HF}, \ref{subfig:Traj_3_HF}]. 

In the spin-flop regime, the first minimum in the low-frequency mode dispersion corresponds to the transition between the AF-coupled phase and the spin-flop phase.  
As was explained previously, in this regime, the static directions of the magnetic moments are continuously altered. The exact angles were shown in Fig. \ref{fig:Eq_angles}. In Fig. \ref{subfig:LF_dyn_energies}, the derivative of the total dynamical energy of the low-frequency mode changes sign at the transition from AF to the spin-flop regime. 
According to Figs.~\ref{subfig:Traj_9_LF},~\ref{subfig:Traj_3_LF} at this point, the OOP components of the magnetizations reach their minimum and increase with the applied field, while for the IP components the same trend continues up to $\approx$ \SI{104}{\milli\tesla}, i. e. the second maximum in the $f(\mathbf{H})$ relation. 
Although the IP components of the magnetization precession dominate, due to the influence of the strong demagnetizing fields, one cannot neglect the OOP components, as they are found to completely determine the behavior of the field-frequency dependence in the spin-flop regime. 

After the second maximum in the field-frequency dependence for the low-frequency mode, the amplitude of the ED-FMR response significantly decreases. On the other hand, the amplitude of the high-frequency mode is increasing after the second maximum. This corresponds to the transition between the in-phase and out-of-phase precessions.
Similarly to the ED-FMR case, the VNA-FMR amplitude for the high-frequency mode also increases after the precession type switching point (\SI{104}{\milli\tesla}). 

After the saturation, in both cases, the high-frequency mode has a higher intensity and Kittel-like behavior is observed for both modes. The OOP dynamic magnetization components are increasing and the IP components are decreasing with the further increase of the external field, slowly approaching a circular precession trajectory.

One can also notice a rapid decrease in the amplitude of the ED-FMR signal, at the points where the derivative of the $f(\mathbf{H})$ dependence becomes zero. This is connected to the experimental technique. Since the field-modulation approach is used, one can measure an output signal using the lock-in technique only, when the system goes in and out of resonance due to the modulation field. On the other hand, if the derivative of the $f(\mathbf{H})$ dependence is close to zero, the modulation field induces no changes in the system behavior, and therefore, no signal is created at the lock-in frequency.

  \begin{figure*}[t]
  	\begin{subfigure}[b]{0.5\textwidth}
  		\captionsetup{labelfont=bf,font=large}
  		\caption{}
  		\includegraphics[width=\linewidth]{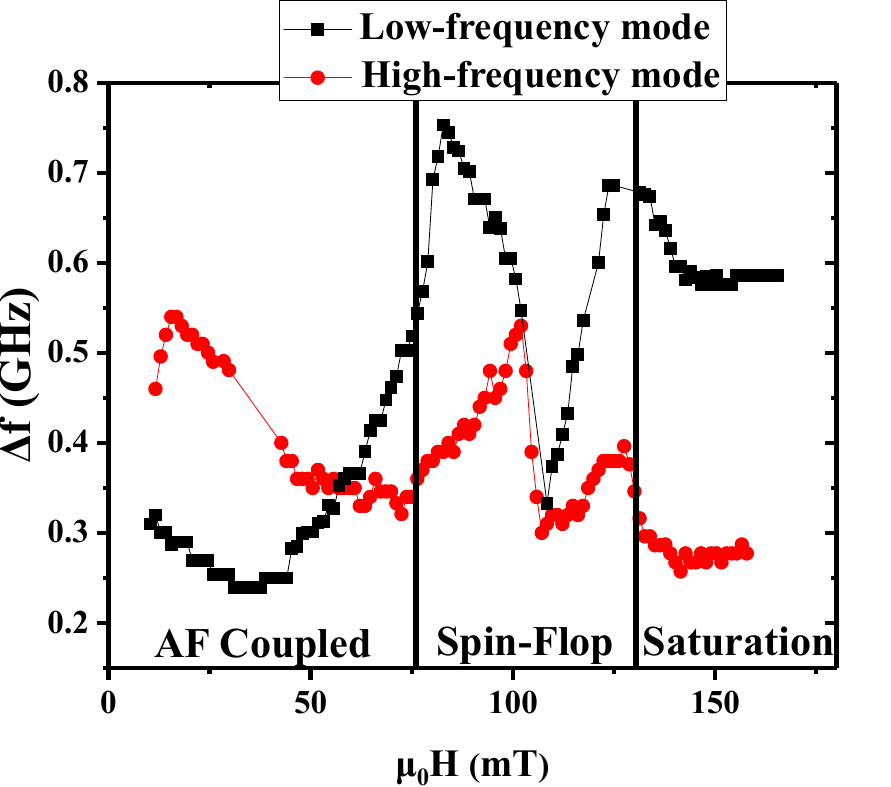}
  		\label{subfig:spin_pump_edfmr}
  	\end{subfigure}
  	\begin{subfigure}[b]{0.5\textwidth}
  		\captionsetup{labelfont=bf,font=large}
  		\caption{}
  		\includegraphics[width=\linewidth]{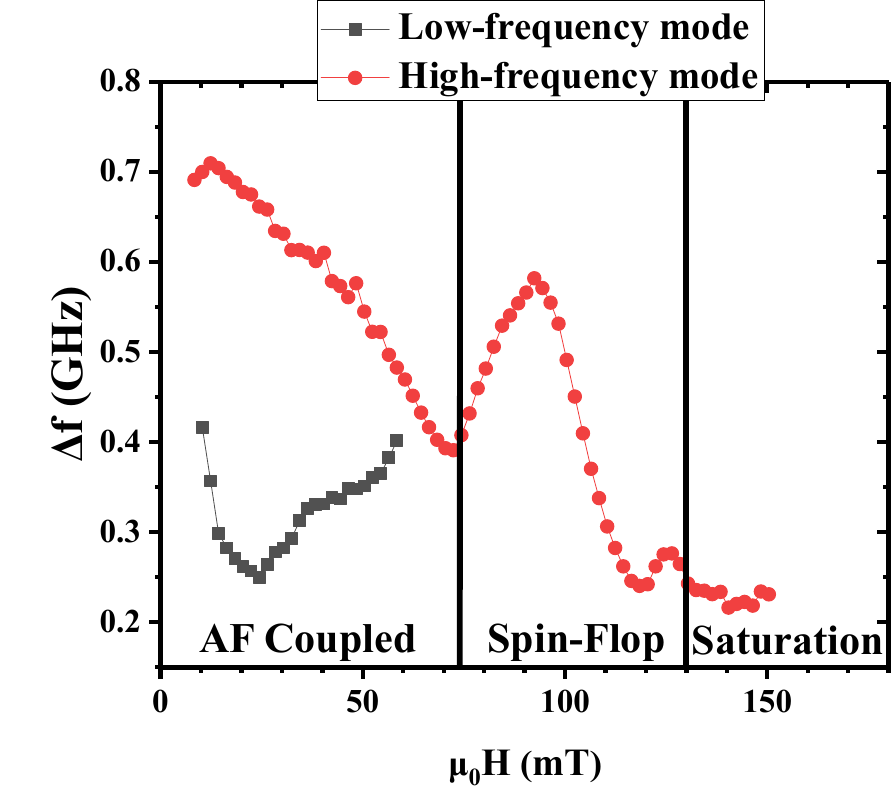}
  		\label{subfig:spin_pump_fmr}
  	\end{subfigure}
  	\caption{Dependence of the linewidth of the 9/3 sample versus the external applied field for low-frequency (in black) and high-frequency (in red) modes for different static configurations, measured using ED-FMR \textbf{(a)} and VNA-FMR \textbf{(b)}. }
  	\vspace{30pt}
  	\label{fig:Spin_pump}
  \end{figure*}
  

\subsection{Effect of the dynamical magnetization on mutual spin-pumping and linewidth}

We can now compare and interpret the linewidth of the resonances for each mode in different field regimes [Fig. \ref{fig:Spin_pump}]. 
Note that only ED-FMR allows for detecting both resonance modes in all the aforementioned regimes. 

\begin{figure*}[t]
	\begin{subfigure}[b]{0.5\textwidth}
		\captionsetup{labelfont=bf,font=large}
		\caption{}
		\includegraphics[width=\linewidth]{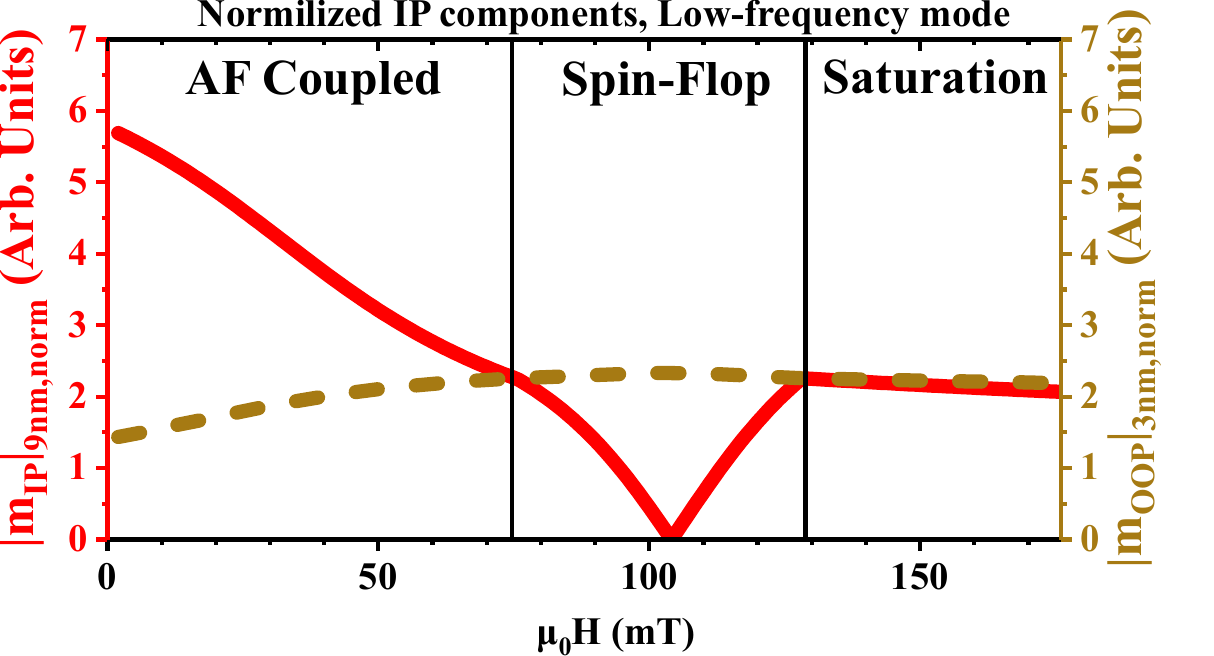}
		\label{subfig:Traj_LF_norm}
	\end{subfigure}
	\begin{subfigure}[b]{0.5\textwidth}
		\captionsetup{labelfont=bf,font=large}
		\caption{}
		\includegraphics[width=\linewidth]{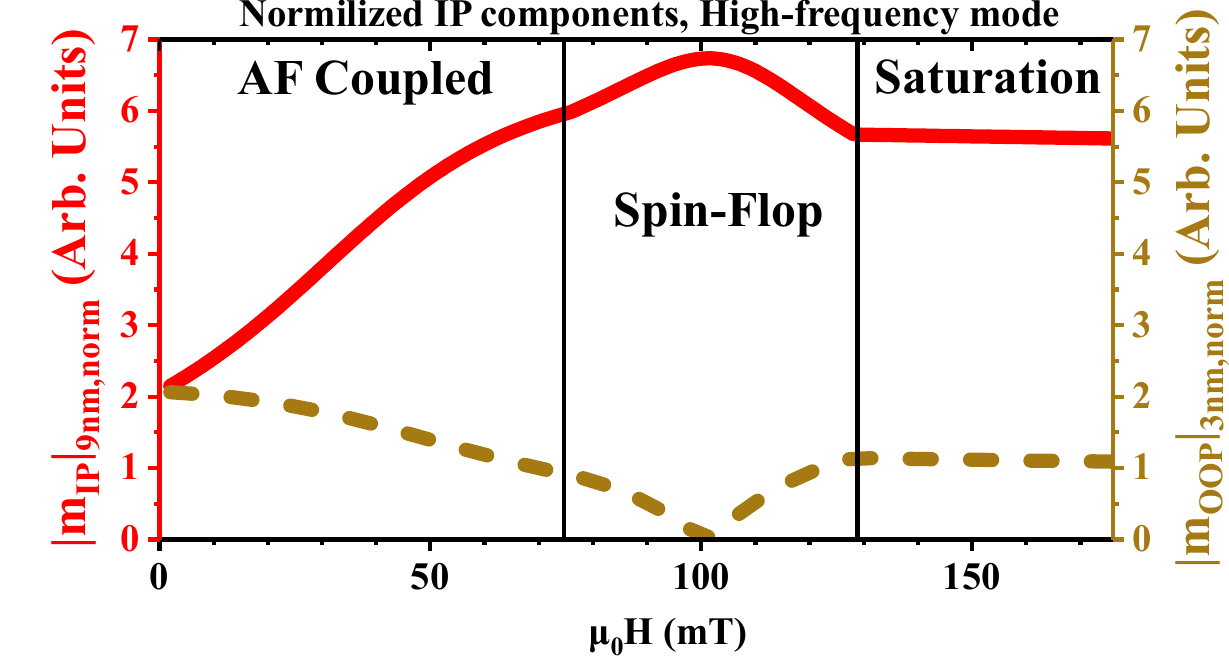}
		\label{subfig:Traj_HF_norm}
	\end{subfigure}
	\caption{Magnitudes of the in-plane dynamical components of the magnetizations normalized by thicknesses and magnetization values of the 9 nm and 3 nm Py layers, for the low-frequency \textbf{(a)} and the high-frequency mode \textbf{(b)}. }
	\vspace{30pt}
	\label{fig:Trajectories_norm}
\end{figure*}

It was shown in Refs. \cite{Takanashi} and \cite{Yang_2016} that in the saturated regime, for symmetric SAFs, the mode that corresponds to the out-of-phase precession (low-frequency mode in our case) has a significantly higher linewidth than the mode corresponding to the in-phase precession (high-frequency mode in our case). 
This linewidth difference is connected to the exchange of the spin-angular momentum (\emph{mutual spin-pumping}) between the layers. 
In the case of in-phase precession, the angular momentum that is leaking from one of the layers due to spin-pumping, is being compensated by the spin-current of the same sign but opposite direction from the other layer. 
When the precessions are out-of-phase, the mutual spin-pumping leads to an additional damping-like torque, increasing the linewidth. 

As can be seen in Fig. \ref{subfig:spin_pump_edfmr}, we observe an increase in linewidth for the low-frequency mode (black squares) in the saturated regime, similar to Ref. \cite{Yang_2016}.
On the other hand, in the AF-coupled regime, the situation is reversed, with a lower linewidth being observed for the low-frequency mode.
The high-frequency mode (red dots) exhibits a larger linewidth in the AF-coupled regime than we see in the saturated regime. This effect is related to the change in the precession type when going from the saturation to the AF-coupled regime.
Additionally, in the AF-coupled regime the difference in linewidth between both modes does not remain constant, as it does in the saturated regime. 
Compared to the VNA-FMR results, Fig. \ref{subfig:spin_pump_fmr}, the same linewidth trend can be observed. However, we were unable to observe the low frequency mode in the spin-flop and saturation regime for VNA-FMR measurements and, therefore, cannot interpret the linewidth in these regimes.
The qualitative behavior of the high-frequency mode is the same for both ED-FMR and VNA-FMR measurements.

In the AF-coupled regime, at low fields the linewidth, after a small increase for the high-frequency mode, is decreasing for both modes. 
After $\approx$ \SI{37}{\milli\tesla}, the linewidth for the low-frequency mode starts to increase and the difference in linewidth between the two modes gradually decreases. 

In order to understand this behavior using the effect of mutual spin-pumping in the context of dynamical components, shown in Fig. \ref{fig:Trajectories}, one has to remember that the MR response is proportional to the precession angle, which is in turn in our configuration proportional to the IP dynamical magnetization component [Eq. (\ref{eq:angletoMip})].
The strength of the spin-current created by the spin-pumping from one layer into the other can be written as \cite{Tserkovnyak_spinpumping}:
 \begin{equation}
 \vec{j_s}_{i} = \frac{\hbar g^{\uparrow \downarrow}}{4\pi M_s^2}\left[\mathbf{M}_i(t)\times \frac{d\mathbf{M}_i(t)}{dt}\right] = \frac{\hbar g^{\uparrow \downarrow}}{4\pi}\mathbf{m}_i(t),
 \label{eq:spin-pumping}
 \end{equation}
 where $g^{\uparrow \downarrow}$ is the spin-mixing conductance of the ferromagnet/normal metal interface and $\mathbf{M}_i(t)$ are the magnetizations of the ferromagnetic layers and $\mathbf{m}_i$ are the dynamical components of the magnetizations of each layer. 
 
Although Eq. (\ref{eq:spin-pumping}) is written in terms of  $\mathbf{m}_i$, we believe that, when talking about mutual spin-pumping, one has to compare dynamical magnetic moments, and not magnetizations, since the former directly corresponds to the transferred angular momentum. 
In case of the same precession angle, the layer with the higher total magnetic moment will also pump a higher amount of angular momentum into the adjacent layer.
Therefore, in the analysis of the mutual spin-pumping, to directly compare IP and OOP components for the \SI{3}{\nano\meter} and \SI{9}{\nano\meter} layers, one has to increase the weight of the components in case of the \SI{9}{\nano\meter} layer proportional to the thickness ratio (3 times) and magnetizations ratio [see Table \ref{table:SQUID}].
The weighted data $|m_\mathrm{IP}|_\mathrm{9nm,norm}$ (red) and $|m_\mathrm{IP}|_\mathrm{3nm,norm}$ (blue) is plotted for the low frequency mode in Fig. \ref{subfig:Traj_LF_norm} and for the high-frequency mode in Fig. \ref{subfig:Traj_HF_norm}.

In case of the low-frequency mode, Fig. \ref{subfig:Traj_LF_norm}, in the AF-coupled regime the larger difference between $|m_\mathrm{IP}|$ for the \SI{9}{\nano\meter} and \SI{3}{\nano\meter} layers leads to an overall anti-damping-like torque and lower linewidth [Fig. \ref{subfig:spin_pump_edfmr}]. 
As the asymmetry in the pumped angular momentum decreases, the linewidth increases. 
Close to the spin-flop regime, the difference in the values of $|m_\mathrm{IP}|$ is minimal and the linewidth is maximal.
For the high-frequency mode in the same regime, Fig. \ref{subfig:Traj_HF_norm}, one can see a negligible difference in $|m_\mathrm{IP}|$ at \SI{0}{\milli\tesla}  with a maximum linewidth [Fig.  \ref{subfig:spin_pump_edfmr}]. As the asymmetry increases the linewidths decreases with a minimum at \SI{76}{\milli\tesla}.

Upon entering the spin-flop regime, not only the dynamical trajectories, but also the static directions of the magnetizations are changing. 
Changes in the static directions of the magnetizations have direct influence on whether the pumped spin-current will act as damping- or field-like torque to the second layer. 
When the layers are at $90^\circ$ with respect to each other ($\approx$ \SI{104}{\milli\tesla}), the dynamical component of one layer is exactly parallel to the magnetization orientation of the other one. 
Moreover, according to the Figs. \ref{fig:Trajectories} and \ref{fig:Trajectories_norm}, at this point, the precession trajectory of one of the layers (9 nm for low-frequency mode and 3 nm for high-frequency mode) vanishes and the SAF essentially behaves as a single magnetic layer. Therefore no mutual spin-pumping occurs and the linewidths for both modes are equal.
This value of the linewidth also lies in between the maximum and minimum observed for both modes. 
This supports the fact that at this point, spin-pumping from one of the layers has no influence on the damping of the second layer and vice versa. 

Finally, entering the saturation regime, the low-frequency mode corresponds to the out-of-phase precession and, therefore, its linewidth increases up to a stable value. 
Concurrently, the high-frequency mode now corresponds to in-phase precession and its linewidth becomes lower in saturation, in agreement to the results obtained in \cite{Yang_2016}. 

\section{Conclusions}
The dynamics of asymmetric synthetic antiferromagnets has been studied using two complementary experimental techniques, namely ED-FMR and VNA-FMR. 
Both dynamical modes of the system were detected in all the possible relative static configuration regimes (AF-coupled, spin-flop and saturation). 
It was shown that the behavior of the frequency-field dependence in the AF-coupled regime is governed by the dynamic interlayer exchange coupling and Zeeman interactions. 
The obtained results are in agreement with the modeling based on the coupled LL equations as well as with static resistance and magnetometry measurements. 
We explained the difference in the ED-FMR and VNA-FMR output signal intensities based on the magnitudes of the dynamic magnetization orbits. 

The linewidths of two different modes for all static configurations were compared. 
The differences in linewidth for low- and high-frequency modes reverse its sign across the spin-flop regime. 
The linewidth gap between the two modes decreases as the external field approaches the critical value of the transition to the spin-flop regime. 
In saturation, the linewidth difference is stable, in accordance with a previous study \cite{Yang_2016}. 
Both results can be explained considering mutual spin-pumping between the layers, in combination with a change in precession phase (in-phase or out-of-phase). 
The results show that mutual spin-pumping in the AF-coupled regime opens the possibility to tune the linewidth in active devices based on SAFs with the application of a small external field, in contrast to the saturation regime. 

\section{Acknowledgments}
S. S., C. F., A. T., G. A. and A. M. D. acknowledge funding from the European Union’s Horizon 2020 research and innovation programme under grant agreement No. 737038 (TRANSPIRE). R. G. acknowledges financial support from Fondecyt Iniciaci\'{o}n, Grant 11170736 and Basal Program for Centers of Excellence, Grant FB0807 CEDENNA, CONICYT. Support of both the Nanofabrication and Structural Characterization Facilities Rossendorf at the Ion Beam Center is gratefully acknowledged.

\bibliographystyle{abbrv}
\bibliography{SAFS}

\begin{thebibliography}{10}
\bibitem{Endnote_1}
One has to note that the VNA-FMR signal is detectable through the whole AF
  coupled regime and the colorplot is a representation of the relative
  amplitudes.
\end{thebibliography}

\end{document}